\documentstyle[prl,aps,floats,epsf]{revtex}

\begin{document}
\draft

\twocolumn[\hsize\textwidth\columnwidth\hsize\csname @twocolumnfalse\endcsname
\title{Reply to Comment on ``Quantum Phase Transition of
  Randomly-Diluted Heisenberg Antiferromagnet
  on a Square Lattice''}
\author{Synge Todo$^1$\cite{Todo}, Hajime Takayama$^1$, and Naoki
Kawashima$^2$}
\address{${}^1$Institute for Solid State Physics, University of Tokyo,
  Kashiwa 277-8581, Japan}
\address{${}^2$Department of Physics, Tokyo Metropolitan University,
  Tokyo 192-0397, Japan}
\date{\today}
\maketitle
\vspace*{2em}
]
\narrowtext

In Ref.~\cite{KatoTHKMT}, we have studied the quantum phase transition
of the two-dimensional diluted Heisenberg antiferromagnet by using the
quantum Monte Carlo method.  The two main conclusions in the article are
i) the critical concentration of magnetic sites is equal to the
classical percolation threshold ($p_{\rm cl} = 0.592746$), and ii) the
value of the critical exponents, such as $\beta$ and $\Psi$,
significantly depends on the spin size $S$.  In \cite{Sandvik}, Sandvik
has presented the result of his quantum Monte Carlo simulation for the
same system only in the case of $S=\frac{1}{2}$, in which the
temperature dependence of the static structure factor is studied quite
precisely.  Based on his data, he claimed that the data in
Ref.~\cite{KatoTHKMT} should be seriously affected by the finiteness of
temperature in our simulation.  In addition, he calculated the cluster
magnetization $m_{\rm cl}(L)$ just at $p=p_{\rm cl}$, and claimed that
it could remain finite even in the thermodynamic limit in the same way
as the classical ($S=\infty$) case, although quite large finite-size
corrections to its asymptotic behavior should be considered in the
quantum case.  As we will discuss below, however, his data do not
conflict with our data nor with our conclusions.  In particular, the
data presented in Fig.~1(b) in \cite{Sandvik} can be explained from our
view point assuming non-magnetic ground states.


As for his first claim on the effect of the temperature, we agree that
there may be a small underestimation for the zero-temperature values of
the static structure factor.  However, even if we replace our data point
in Fig.~3 in \cite{KatoTHKMT} for $L=48$ by what we can expect from
Fig.~1(a) in \cite{Sandvik}, it would result in only a minor change in
the estimate of $\Psi$.  Therefore, the error caused by this
underestimation, if any, does not seem to be large enough to invalidate
our conclusions.

In addition, in Ref.~\cite{KatoTHKMT}, we have presented the result of
another finite-size scaling analysis (Fig.~4), in which the data at
higher temperatures are also used together with the one at the lowest
temperature in the simulation.  In this analysis, we have explicitly
taken into account the finite-temperature effect on the static structure
factor by introducing the dynamical exponent $z$.  For $S=1$, the value
of the exponent $\Psi$ coincides with that obtained by the former
analysis.  On the other hand, for $S=\frac{1}{2}$, it becomes slightly
larger (about 9\%) than the other estimate based on Fig.~3 in
\cite{KatoTHKMT}.  This small difference may be due to the
finite-temperature effect pointed out in \cite{Sandvik}, though we have
not commented on it in \cite{KatoTHKMT}.  However, even the estimate
based on the finite-temperature analysis significantly differs from
those for $S\ge1$.  Therefore, we think our second conclusion on the
non-universal behavior can not be excluded.

\begin{figure}[t]
 \null\vskip2mm
 \centerline{\epsfxsize=0.30\textwidth\epsfbox{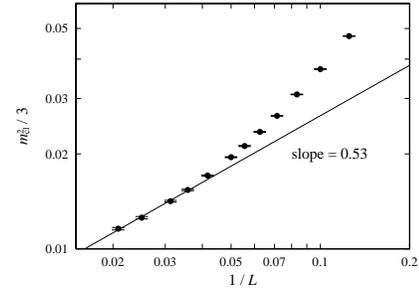}}
 \caption{Log-log plot of cluster magnetization
 (Ref.\protect\cite{Sandvik}).  The solid line is obtained by
 least-squares fitting for $L\ge28$.}
 \label{fig:log-log}
\end{figure}

It was also suggested~\cite{Sandvik}, based on the calculation of the
magnetization of the largest connected cluster for each sample, that all
the critical exponents, including $\Psi$ and $\beta$, are equal to the
classical values.  The data were plotted against $x\equiv 1/L^{D/2}$,
and appeared to suggest a finite value in the thermodynamic limit.  From
our point of view, however, his data can be
interpreted as follows.  The quantity $m_{\rm cl}^2(L)$ defined in
Ref.~\cite{Sandvik} should be asymptotically scaled as
\begin{equation}
 m_{\rm cl}^2(L) \sim S_s(L,0,p_{\rm cl}) / L^{2D-d},
\end{equation}
where $S_s(L,T,p)$ is the static structure factor (Eq.~(4) in
Ref.~\cite{KatoTHKMT}) and $L^D$ is the average number of spins in the
largest connected clusters in a system of linear size $L$.
According to our results~\cite{KatoTHKMT}, on the other hand,
$S_s(L,0,p_{\rm cl})$ diverges as $L^{2D-d-\alpha}$ with $\alpha=0.52$
for $S=\frac{1}{2}$, and consequently $m_{\rm cl}^2(L)$ should be scaled
as $L^{-0.52}$.
In Fig.~\ref{fig:log-log}, we show a log-log plot of the cluster
magnetization data by Sandvik~\cite{Sandvik}.  One sees that the data
for $L\ge28$ clearly scales as $L^{-0.53}$, which is fully consistent
with our results mentioned above.

The authors are grateful to A.~W. Sandvik for allowing us to use his
numerical data.

\null

\end{document}